\documentstyle[aps, epsfig]{revtex}

\def\xslash#1{{\rlap{$#1$}/}}
\def\half{\frac{1}{2}}
\def\beq{\begin{equation}}
\def\eeq{\end{equation}}
\def\beqa{\begin{eqnarray}}
\def\eeqa{\end{eqnarray}}
\def\iar{\begin{array}{l}}
\def\ear{\end{array}}

\begin{document}
\draft
\title{Renormalization of the Cabibbo-Kobayashi-Maskawa Matrix in Standard Model}
\author{Yong Zhou$^{a}$}
\address{$^a$ Institute of Theoretical Physics, Chinese Academy of Sciences, 
         P.O. Box 2735, Beijing 100080, China}
\maketitle

\begin{abstract}
We have investigated the present renormalization prescriptions of 
Cabibbo-Kobayashi-Maskawa (CKM) matrix, and found there is still not an 
integrated prescription to all loop levels. In this paper We propose a new 
prescription which is designed for all loop levels in the present 
perturbative theory. This new prescription will keep the unitarity of the bare 
CKM matrix and make the amplitude of an arbitrary physical process involving quark
mixing convergent and gauge independent. 
\end{abstract}

%11.10.Gh: Renormalization
%12.15.Lk: Electroweak radiative corrections 
%12.15.Hh: Determination of Kobayashi-Maskawa matrix elements

Since the exact examination of the Cabibbo-Kobayashi-Maskawa (CKM) quark mixing 
matrix \cite{c1} has been developed quickly, the renormalization of CKM matrix 
becomes very important \cite{c2}. This was realized for the Cabibbo angle with 
two fermion generations by Marciano and Sirlin \cite{c3} and for the CKM matrix 
of the three-generation SM by Denner and Sack \cite{c4} more than a decade ago. 
In recent years many people have discussed this issue \cite{c5}, but a completely
self-consistent prescription to all loop levels has been not obtained. In this 
paper we will try to solve this problem.

In general, a CKM matrix renormalization prescription should satisfy the three 
criterions \cite{c6}:

\begin{enumerate}
\item In order to keep the transition amplitude of any physical process involving
quark mixing ultraviolet finite, the CKM counterterm must cancel out the 
ultraviolet divergence left in the loop-corrected amplitudes. 
\item It must guarantee such transition amplitude gauge parameter independent 
\cite{c7}, which is a fundamental requirement. 
\item SM requires the bare CKM matrix $V^0$ is unitary,

\beq
\sum_k V^0_{ik} V^{0\ast}_{jk}\,=\,\delta_{ij}
\eeq
with $i,j,k$ the generation index and $\delta_{ij}$ the unit matrix element. 
If we split the bare CKM matrix element into the renormalized one and its 
counterterm 

\beq
  V^0_{ij}=V_{ij}+\delta V_{ij}
\eeq
and keep the unitarity of the renormalized CKM matrix, the unitarity of the bare 
CKM matrix requires 

\beq
  \sum_k (\delta V_{ik}V^{\ast}_{jk}+V_{ik}\delta V^{\ast}_{jk}+
  \delta V_{ik}\delta V^{\ast}_{jk})\,=\,0
\eeq
\end{enumerate}

Until now there are many papers discussing this problem. The modified minimal 
subtraction ($\overline{MS}$) prescription \cite{c8} is the simplest one, but 
the introduced $\mu^2$-dependent term is very complicated to be dealt with. In 
the on-shell renormalization framework, however, there is still not an integrated
CKM renormalization prescription. The early prescription which used the $SU_L(2)$
symmetry of SM to relate the CKM counterterm with the fermion wave-function 
renormalization constants (WRC) \cite{c4,c9} is a delicate and simple 
prescription, but unfortunately it reduces the physical amplitude involving 
quark mixing gauge dependent\footnote{It is easy to be understood since the 
$SU_L(2)$ symmetry of SM has been broken by the Higgs mechanism} \cite{c10,c11}. 
A revised version of this prescription is to replace the on-shell fermion WRC 
in the CKM counterterm with the one calculated at zero momentum \cite{c10}. 
Another revised method is to rearrange the off-diagonal quark WRC in a manner 
similar to the pinch technique \cite{c11}. Different from the idea of 
Ref.\cite{c4}, another idea is to formulate the CKM renormalization 
prescription with reference to the case of no mixing of fermion generations.
The main idea of this prescription is to make the transition amplitude of W gauge
boson decaying into up-type and down-type quarks equal to the same amplitude but 
in the case of no mixing of quark generations \cite{c12}. All of the mentioned 
prescriptions are only applied to one-loop level and are very difficult to be 
generalized to higher loop levels \cite{c10,c11,c12}. So we want to propose a new
method to solve this problem. 

We will renormalize the CKM matrix through two steps. First we introduce a CKM
counterterm which makes the physical amplitude of $W^{+}\rightarrow u_i \bar{d}_j$
ultraviolet convergent and gauge independent. Next we mend it to satisfy the
unitary condition of Eq.(3), simultaneously keep the divergent and
gauge-dependent part of it unchanged. In order to elaborate our idea clearly we
firstly introduce the n-loop ($n\geq 1$) decaying amplitude of 
$W^{+}\rightarrow u_i \bar{d}_j$ as follows (here all of the counterterms lower 
than n-loop level merge into the formfactors):

\beq
  T_n\,=\,A_L[F_{Ln}+V_{ij}(\frac{\delta g_n}{g}+\half\delta Z_{Wn})+
  \half\delta \bar{Z}^{uLn}_{ik}V_{kj}+\half V_{ik}\delta Z^{dLn}_{kj}+
  \delta V_{ijn}]+A_R F_{Rn}+B_L G_{Ln}+B_R G_{Rn}
\eeq
with $g$ and $\delta g$ the $SU(2)$ coupling constant and its counterterm, 
$\delta Z_W$ the W boson WRC, $\delta\bar{Z}^{uL}$ and $\delta Z^{dL}$ the 
left-handed up-type and down-type quark's WRC \cite{c13}. The added denotation 
"n" represents the n-loop result, and

\beqa \iar
  A_L\,=\,\frac{g}{\sqrt{2}}\bar{u}_i(p_1){\xslash \varepsilon}\gamma_L 
  \nu_j(q-p_1)\,, \\
  B_L\,=\,\frac{g}{\sqrt{2}}\bar{u}_i(p_1)\frac{\varepsilon\cdot p_1}{M_W}
  \gamma_L \nu_j(q-p_1)\,.
\ear \eeqa
with $\varepsilon^{\mu}$ the W boson polarization vector, $\gamma_L$ and 
$\gamma_R$ the left-handed and right-handed chiral operators, $M_W$ the W boson
mass. Similarly, replacing $\gamma_L$ with $\gamma_R$ in above equations we 
get $A_R$ and $B_R$ respectively. $F_{L,R}$ and $G_{L,R}$ are four formfactors. 
Here we only care about the coefficient of $A_L$ which contains the n-loop CKM
counterterm. The simplest method to make the amplitude $T_n$ convergent and gauge
independent is to make the coefficient of $A_L$ equal to zero (the remaining 
terms should be convergent and gauge independent \cite{c10}). That's to say

\beq
  \delta V_{ijn}\,=\,-F_{Ln}-V_{ij}(\frac{\delta g_n}{g}+\half\delta Z_{Wn})-
  \half\delta \bar{Z}^{uLn}_{ik}V_{kj}-\half V_{ik}\delta Z^{dLn}_{kj}
\eeq

Obviously such CKM counterterm cannot guarantee the unitarity of the bare CKM 
matrix. It needs to be mended. Here we introduce a new denotation 
$\delta\bar{V}_n$ to denote the amended CKM counterterms which will satisfy the 
unitary condition of Eq.(3) to n-loop level. Our method is to by recursion 
construct $\delta\bar{V}_n$ through 
$\delta V_n,\delta\bar{V}_{n-1},\cdot\cdot\cdot\cdot,\delta\bar{V}_1$. Here we 
state that $\delta V_n$ is obtained by using 
$\delta\bar{V}_{n-1},\cdot\cdot\cdot\cdot,\delta\bar{V}_1$ as the lower loop CKM
counterterms in Eq.(6). Now the unitary condition of Eq.(3) becomes 

\beqa \iar
  \delta\bar{V}_1 V^{\dagger}+V \delta\bar{V}^{\dagger}_1\,=\,0\,,\\
  \delta\bar{V}_2 V^{\dagger}+V \delta\bar{V}^{\dagger}_2\,=\,-
  \delta\bar{V}_1 \delta\bar{V}^{\dagger}_1\,,\\
  \delta\bar{V}_3 V^{\dagger}+V \delta\bar{V}^{\dagger}_3\,=\,-
  \delta\bar{V}_1 \delta\bar{V}^{\dagger}_2-
  \delta\bar{V}_2 \delta\bar{V}^{\dagger}_1\,,\\
  \cdot\cdot\cdot\cdot\cdot\cdot \\
  \delta\bar{V}_n V^{\dagger}+V \delta\bar{V}^{\dagger}_n\,=\,-
  \delta\bar{V}_1 \delta\bar{V}^{\dagger}_{n-1}-
  \delta\bar{V}_2 \delta\bar{V}^{\dagger}_{n-2}\cdot\cdot\cdot-
  \delta\bar{V}_{n-2}\delta\bar{V}^{\dagger}_2-
  \delta\bar{V}_{n-1}\delta\bar{V}^{\dagger}_1\,,\\
  \cdot\cdot\cdot\cdot\cdot\cdot 
\ear \eeqa
In order to solve these equations, we introduce a set of symbols $B_n$

\beqa \iar
  B_0\,=\,0\,, \\
  B_n\,=\,\sum^{n-1}_{i=1} -\delta\bar{V}_i \delta\bar{V}^{\dagger}_{n-i}\,.
\ear \eeqa
Obviously $B_n$ satisfies

\beq
  B_n\,=\,B^{\dagger}_n
\eeq
Assuming that we have obtained the counterterms 
$\delta\bar{V}_1,\delta\bar{V}_2,\cdot\cdot\cdot\cdot,\delta\bar{V}_{n-1}$ and 
$\delta V_n$, the n-loop CKM counterterm $\delta\bar{V}_n$ is determined as 
follows:

\beq
  \delta\bar{V}_n\,=\,\half(\delta V_n -V\delta V^{\dagger}_n V+B_n V)
\eeq
At one-loop level this result is similar as Eq.(11) of ref.\cite{c6}.
It is very easy to see that such CKM counterterm satisfies Eqs.(7) to n-loop 
level.

The remaining problem is to test whether the amended CKM counterterm 
$\delta\bar{V}_n$ has the same divergent and gauge-dependent terms as 
$\delta V_n$, which is the requirement of making the physical amplitude involving
quark mixing ultraviolet finite and gauge independent. Our answer is positive
to this question. Based on the renormalizability and predictability of SM, we 
can predict that the divergent and gauge-dependent part of $\delta V_n$ must 
satisfy the unitary condition of Eq.(3) at n-loop level

\beq
  \delta V^{DG}_n V^{\dagger}+V\delta V^{DG\dagger}_n\,=\,B^{DG}_n
\eeq
where the superscript "DG" denotes the divergent/gauge-dependent part of 
the quantity. This is because if not so the unitary condition of Eq.(3) will 
require the divergent and gauge-dependent part of the CKM counterterm different
from $\delta V_n$ thus will reduce the physical amplitude of 
$W^{+}\rightarrow u_i \bar{d}_j$ divergent and gauge dependent. At one-loop level
this relationship has been proven in actual calculations \cite{c6,c10}. So from 
Eq.(10) and (11) it is obtained

\beq
  (\delta\bar{V}^{DG}_n-\delta V^{DG}_n)V^{\dagger}\,=\,\half(B^{DG}_n-
  \delta V^{DG}_n V^{\dagger}-V\delta V^{DG\dagger}_n)\,=\,0
\eeq
This identity manifests that
\beq
  \delta\bar{V}^{DG}_n\,=\,\delta V^{DG}_n
\eeq

Now we have obtained the proper CKM counterterms to all loop levels, which satisfy
the unitary condition of the bare CKM matrix and make the physical amplitude 
involving quark mixing convergent and gauge independent. We state that at present 
all of the CKM renormalization prescriptions are only applied to one-loop level 
and an integrated prescription to all loop levels doesn't appear. This situation
shows that the renormalization of CKM matrix isn't an easy job. Our prescription 
affords a more straightforward and simple method to solve this problem and it will
be easy to calculate the CKM counterterm in actual calculations (the n-loop CKM 
counterterm is shown in Eq.(10) and the explicit one-loop result is shown in 
appendix). On the other hand we suppose our prescription will not break the 
present symmetries of SM, e.g. Ward-Takahashi identity, because it only changes
the values of CKM matrix elements from $V^0_{ij}$ to $V_{ij}+\delta\bar{V}_{ij}$. 

\vspace{5mm} {\bf \Large Acknowledgments} \vspace{2mm} 

The author thanks professor Xiao-Yuan Li for his useful guidance and Dr. Hu 
qingyuan for his sincerely help (in my life).

\vspace{5mm} {\bf \Large Appendix} \vspace{2mm} 

In this appendix we give the explicit result of $\delta\bar{V}_1$. In the on-shell
renormalization framework $\delta g$ and $\delta Z_W$ are both real quantities, so
we can obtain the following result through Eq.(10), (8) and (6)

\beqa \iar
  \delta\bar{V}_{ij1}\,=\,\half(\delta V_{ij1}-\sum_{kl}V_{ik}
  \delta V_{kl1}^{\dagger}V_{lj}) \\ \hspace{9mm} \,=\,
  \half(\sum_{kl}V_{ik}F_{L1lk}^{\ast}V_{lj}-F_{L1ij})+
  \frac{1}{4}\sum_{k}(\delta\bar{Z}^{uL1\ast}_{ki}-\delta\bar{Z}^{uL1}_{ik})V_{kj}
  +\frac{1}{4}\sum_{k}V_{ik}(\delta Z^{dL1\ast}_{jk}-\delta Z^{dL1}_{kj})
\ear \eeqa
which is gauge independent since $\delta V_{ij1}$ is gauge independent \cite{c10}.
The ultraviolet divergence of $\delta\bar{V}_{ij1}$ is

\beqa \iar
  \delta\bar{V}_{ij1}|_{UV}\,=\,\frac{3\alpha\Delta}{64\pi M^2_W s^2_W}
  [-\frac{2\sum_{k,l\not=j}m_{d,j}m^2_{u,k}V_{il}V^{\ast}_{kl}V_{kj}}{m_{d,l}-
  m_{d,j}}+\frac{2\sum_{k,l}m_{d,j}m^2_{u,k}V_{il}V^{\ast}_{kl}V_{kj}}{m_{d,l}+
  m_{d,j}}-
  \frac{2\sum_{k\not=i,l}m_{u,i}m^2_{d,l}V_{il}V^{\ast}_{kl}V_{kj}}{m_{u,k}-
  m_{u,i}}+ \\ \hspace{19mm}
  \frac{2\sum_{k,l}m_{u,i}m^2_{d,l}V_{il}V^{\ast}_{kl}V_{kj}}{m_{u,k}+
  m_{u,i}}+V_{ij}(\sum_k V_{ik}V^{\ast}_{ik}m^2_{d,k}+
  \sum_k V_{kj}V^{\ast}_{kj}m^2_{u,k}-2 m^2_{d,j}-2 m^2_{u,i})]\,.
\ear \eeqa
where $\alpha$ is the fine structure constant, $M_W$ is the W boson mass, $s_W$ is
the sine of the weak mixing angle $\theta_W$, and 
$\Delta=2/(D-4)+\gamma_E-\ln(4\pi)+\ln(M^2_W/ \mu^2)$ (D is the space-time 
dimensionality, $\gamma_E$ is the Euler's constant and $\mu$ is an arbitrary 
mass parameter). $m_{u,i}, m_{d,j}$ etc. are up-type and down-type quark's masses.  
The $R_{\xi}$-gauge and the Dimensional regularization have been used. This result
agrees with the results of Ref.\cite{c4} and \cite{c10}.

\end{document}